# NotiMind: Utilizing Responses to Smart Phone Notifications as Affective sensors

Eiman Kanjo[1], Daria J. Kuss[2], and Chee Siang Ang[3]

*[1]Computing and Technology Department, Nottingham Trent University,*
*eiman.kanjo@ ntu.ac.uk*
*[2]Psychology Department , Nottingham Trent University*
*Daria.Kuss@ ntu.ac.uk*
*University of Kent*
*[3]csa8@kent.ac.uk.*

**Abstract**—Today's mobile phone users are faced with large numbers of notifications on social media, ranging from new followers on Twitter and emails to messages received from WhatsApp and Facebook. These digital alerts continuously disrupt activities through instant calls for attention. This paper examines closely the way everyday users interact with notifications and their impact on users' emotion. Fifty users were recruited to download our application NotiMind and use it over a five-week period. Users' phones collected thousands of social and system notifications along with affect data collected via self-reported PANAS tests three times a day. Results showed a noticeable correlation between positive affective measures and keyboard activities. When large numbers of Post and Remove notifications occur, a corresponding increase in negative affective measures is detected. Our predictive model has achieved a good accuracy level using three different classifiers "in the wild" (F-measure 74-78% within-subject model, 72-76% global model). Our findings show that it is possible to automatically predict when people are experiencing positive, neutral or negative affective states based on interactions with notifications. We also show how our findings open the door to a wide range of applications in relation to emotion awareness on social and mobile communication.

**Keywords**— Mobile Sensing, Affective Computing, Mobile Computing, Mobile Social media, Machine Learning

## 1 INTRODUCTION

A notification is generally defined as a visual cue, auditory signal, or haptic alert generated by an application or service to capture the user's attention (Iqbal and Horvitz, 2010). Currently, all major smartphone platforms (i.e., iOS and Android) offer a standardized user interface mechanism for notifications, displaying all notifications in the notification bar located at the top of the screen. Given the increase of mobile applications (apps) on smartphones, notifications are becoming ubiquitous, providing a broad range of information, from system (e. g., app updates) to social information (e. g., a message from a friend). More specifically, notifications on smartphones inform users about a variety of events, such as the arrival of a text message or emails, an incoming phone call, a new comment on one of their social network posts, game-related status updates, system status or the availability of an application update. Notifications hence provide a means for app publishers and advertisers to connect with users.

Within the HCI and Ubicomp research communities, there has been a growing interest in studying how users respond to notifications, with an aim to design better notification delivery systems, which minimize disruption. For instance, it has been found that not all notifications are treated the same way by the users. In response to notifications, users may take immediate action or ignore notifications depending on the importance of a notification as well as the user's current context. It is important to understand how users respond to notifications in the context of smartphones. In contrast to notifications on desktop computers, notifications on smartphones are less likely to be delivered when the users are actively interacting with the device: despite the large amount of time users spend with their smartphones, the device is often not in active use when notifications arrive. Therefore, with smartphones, notifications have some unique characteristics which need to be considered: i) they are delivered through a standardized mechanism, ii) they inform about a larger variety of events, ranging from social messages to system events, and iii) they are pervasive due to the omnipresent nature of smartphones, which are always with the user (Ichikawa, Chipchase et al. 2005; Wiese, Saponas et al. 2013).

While mobile phones and mobile notifications have enhanced the convenience of our life, their obsessive use may have an adverse impact on mental health and wellbeing. This impact is still under investigation and researchers have started to look at various techniques to understand and diagnose problematic mobile phone use and mobile phone addiction (Billieux, Maurage, Olatz-Lopez, Kuss & Griffiths, 2015).

In this paper, we present our mobile phone application NotiMind which aims to:
1. Monitor mobile notifications and gather interaction data unobtrusively based on the amount of notifications received and delivery pattern of notifications on mobile phones.
2. Model user interaction and reaction to mobile notifications and their impact on affective states based on a machine learning approach.

3. Examine patterns of user interaction with mobile phone notifications, status and screen activities and their correlations with affective states of the users.

Our approach is based on utilising machine learning to recognize affective states of smartphone users to assess the users' emotional states. We set up "in the wild" user studies and gathered various types of notifications data from participants' smartphones. These data were collected using a custom-built application, called "NotiMind" which collects a wide range of social and work notifications, such as screen events (e.g. Screen-On and Screen Unlock), Time/Date, ringtone volume. Additionally, NotiMind collects self-reported affective states of users, using a short version of the Positive and Negative Affective States scale (PANAS; Thompson, 2007). The ultimate aim of the study is to unobtrusively recognize individual smartphone users' positive negative emotions, and their relationship with type and frequency of notifications received on their smartphones.

## 2 BACKGROUND

### 2.1 Notifications

Notifications are a feature on smartphones and other devices to keep users informed and engaged. Notifications can alert users to information regarding a range of subjects, including incoming messages, engagement with their social media posts, and availability of WIFI networks or applications updates, curated nearby places according to their geolocations and email content preview (Iqbal and Horvitz 2010), and are commonly presented on a notification panel on top of the screen. Figure 1 shows types of mobile notifications and their main categories.

In the desktop and mobile environment, notifications have been viewed as means to proactively provide awareness of information while users are attending to a primary task. These alerts arrive in form of a brief text and alerting sound (if the volume is on) or vibrating (on mobile phones and smart watches) to catch end-users' attention. The time taken by users to attend to those notifications often depends on how important the context of the notifications is (e.g., a family emergency may require a more immediate response than a work email), and what contexts the users find themselves in when receiving the notification (e.g., a user is more likely to attend to a notification when they are not currently actively engaged in another activity). A main function of notifications is to allow users to switch between work-related and social apps. Such switches are often driven by the user's own need to forage information as required for the current task, or after being proactively alerted about the arrival of new information.

Previous work has shown that although users are aware of the disruptive effects of notifications, they generally appreciate the awareness that notifications provide (Iqbal & Horvitz 2010; Mark, Voida et al. 2012). Specifically in the smartphone context, users are eager to receive notifications as they keep checking their smartphones frequently (Oulasvirta, Rattenbury et al. 2012). The level of importance of notifications varies depending on the categories of the notifying apps (Oulasvirta, Rattenbury et al. 2012; Pielot, Oliveira et al. 2014).

Furthermore, some mobile social network users hardly show what they feel; therefore their friends cannot sense or react to their emotional states appropriately. This is probably because they are unfamiliar with the expression of emotion or are not aware of their own emotions. A possible solution for this problem is adopting emotion recognition technologies which are being extensively studied by the affective computing research society to determine user emotions.

Shirazi et al. (2014) carried out a large-scale assessment of mobile notifications and found that participants rated notifications from messenger applications as the most important ones, at 4.43 out of 5 (5 being the most important). Notifications from the three other communication categories and from the calendar also received high ratings with averages between 3.66 and 3.45. Notifications from the system clearly received the lowest rating (1.6). The large scale assessment (Shirazi, Henze et al. 2014) concluded that important notifications are about people and events, specifically if they notify about communication with other users, inform about other users' actions, or about real life events. Another factor why notifications are considered unimportant is the frequency with which they are created, and whether they provide information about the phone's internal processes.

Pielot, Oliveira et al. (2014) investigated users' response time to various types of notifications and found a wide range of differences. Specifically, the median response time ranged from 3.5 min for messengers on weekends to 27.7 min for email on weekends. The shortest notification responses were provided for messengers (6.6 and 3.5 min) and social network applications (3.8 and 7.0 min / weekday and weekend day, respectively). It was also found that half of the notifications were viewed within a few minutes, and that the majority were attended to within an hour. When notifications arrived, the screen was off in 69.2% of the cases. The speed at which people attended to notifications indicates that notifications often triggered interaction with the phone. Moreover, given the perceived importance of various categories of apps, most users might not want to disable notifications. However this depends on the notification, such as:

1. **Frequency**: some applications (e.g., social media) have the potential to send a large number of alerts by the minute: for chats, for new posts and tagged photos, etc. The notification content is likely to become less important in a situation where a large number of interruptions occur over a short period of time.
2. **Content**: Many mobile applications use notifications as nothing more than marketing vehicles to remind users to use those respective apps. These applications are trivial in their own right, but a potential constant distraction if not turned off.
3. **Importance**: In the case of a large number of applications sending notifications to the user, the user may begin to prioritise these in order of the most immediately important. For instance, notifications which were previously considered important may be re-evaluated with the introduction of new notifications which may render the notification previously considered important to comparatively less relevant.
4. **Category/Source:** Some notifications grab users' attention more than others merely based on the application source which has generated the notification: e.g. email, WhatsApp, Facebook, Udacity, etc.

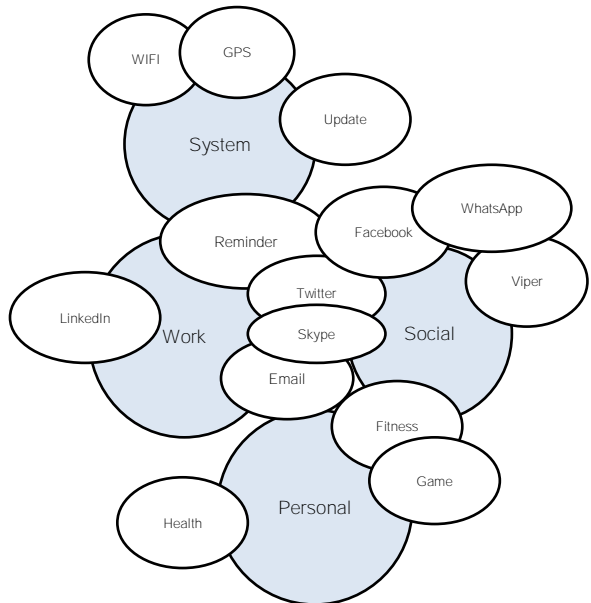

Fig 1. Types of Notifications.

## 2.2 The drawback of notifications

Although notifications serve an important purpose for smartphone users, the number of apps which compete for the user's attention using notifications has grown significantly over the years. Initially designed to raise information awareness, it has been argued that notifications have now become too frequent (Iqbal and Horvitz 2010; Shirazi, Henze et al. 2014), and may be contributing to stress (Westermann et al. 2015).

Extensive research has been conducted investigating interruption of information workers through notifications in a desktop context (Horvitz 2001; Adamczyk & Bailey 2004; Czerwinski, Horvitz et al. 2004). Studies have shown that workers tend to drop their current task to check the notification, and it is difficult for them to return to their previous tasks (Czerwinski, Horvitz et al. 2004). Indeed, a large body of work in HCI has looked into the detrimental effects of digital interruptions. Experiments in a controlled laboratory environment have revealed that notifications arriving at random times are particularly disruptive (Horvitz 2001; Baethge & Rigotti 2013). In field experiments, digital interruptions at work (e.g., email alerts) have been linked to feeling distracted, stressed, and anxious (Kushlev & Dunn 2015).

As discussed previously (Section 2.1), people often attend quickly, if not immediately, to notifications on their smartphones, which arrive at random times, making smartphones particularly disruptive (Adamczyk et al., 2004). Furthermore, due to the omnipresence of smartphones, users are constantly disrupted regardless of where they are, as they are likely to take their phones with them. Even when their phones are set to vibration, studies have demonstrated that people still attend to notifications quickly regardless of the alert type (Pielot, Church et al. 2014; Chang & Tang 2015).

Kushlev, Proulx et al. (2016) provided some evidence that interruptions due to notifications can cause inattention and hyperactivity in the general population. It was found that when people switched on notifications and kept their phones within physical reach, they experienced higher levels of inattention and hyperactivity. They found that notifications draw users' attention away from other ongoing activities, making people more prone to distractions, increasing cognitive load, which may in turn make people experience inattention and hyperactivity. Even when users choose to disable notifications, research has suggested that for some individuals, disabling alerts may produce anxiety over missing out (Pielot and Rello 2015). Such anxiety may lead users to self-interrupt more frequently. In one study, information workers disabled their email notifications for one week; it was found that some workers checked their email even more to avoid missing important emails (Iqbal & Horvitz 2010). Hence, simply turning the notifications off may not be the solution to the disruptive nature of smartphone notifications. This may lead users to compulsively check for missed notifications, for instance from social networking apps due to social pressure (Pielot, Church et al. 2014) and the fear of missing out (FOMO; Przybylski et al. 2013).

Apart from distraction and stress, some studies have suggested that extensive use of smartphones could potentially lead to smartphone addiction (Bayer & Campbell 2012; Billieux et al. 2015; Oulasvirta, Rattenbury et al. 2012). In one study, researchers found that their participants checked their smartphones 34 times a day (Oulasvirta, Rattenbury et al. 2012). Yoon, Lee et al. (2014) investigated the way young people make use of smartphones. It was found that they receive an average of 400 notifications a day. It was claimed that the usage pattern of smartphone by young people bordered on addiction.

However, research on the extent to which mobile phone notifications fare in this is currently limited. Moreover, the traditional clinical approach based questionnaires and interviews typically used in mobile phone addiction research (Lopez-Fernandez, Kuss, et al., 2017) have limitations: health professionals cannot perform continual assessments and interventions for their clients, and the subjectivity of assessments is questionable. This indicates objective data that tracks users in an ecologically valid way is needed in order to better understand mobile phone usage patterns and associated problems.

## 2.3 Machine learning work on notifications

Despite the disruptive nature of notifications, studies have found that users suffered no negative consequences when they were interrupted at opportune times in between work tasks. Therefore, various studies have been conducted by means of machine learning to understand the factors leading to users responding to notifications, with the aim to predict how long it will take for users to respond to a notification, and to understand what the most opportune moments are and what are the best ways to deliver notifications.

Avrahami and Hudson (2006) used machine learning techniques to predict how fast a user responds to an instant message in a desktop computer setting including 16 co-workers at Microsoft with a total of 90,000 messages. Using these data, they trained models with an accuracy of 90.1% to predict if a message would get a response in 0.5, 1, 2, 5, and 10 minutes. Strong predictors of responsiveness were "the amount of user interaction with the system", "the time since the last outgoing message", and "the duration of the current online-status".

Fogarty, Hudson et al. (2005) and Hudson, Fogarty et al. (2003) collected data from four information workers to predict when they were interrupted whilst at work. They concluded that a single microphone, the time of the day, the use of the phone, and the interaction with mouse and keyboard can estimate a worker's interruptibility with an accuracy of 76.3%. Drawing from this, Begole, Matsakis et al. (2004) developed a prototype sensing sounds, motion, using the phone, and using the office door to predict if a worker is potentially available to interruptions.

Rosenthal, Dey et al. (2011) developed a model to predict when a phone should be put in silent mode using experience sampling to collect data on user preferences for different situations. They considered features such as time and location, reason for the alert, and details about the alert (e.g., whether it came from a caller listed in the user's favorites) in their model. An experimental study showed that thirteen out of nineteen participants were satisfied with the accuracy of the automatic muting.

Pielot, Oliveira et al. (2014) identified that features extracted from the phone, such as the user's interaction with the notification center, the screen activity, the proximity sensor, and the ringer mode, can be used to predict how quickly the user will respond to the messages. It was found that with seven high-level features a user's level of attentiveness to mobile messages can be successfully estimated, with an overall accuracy of 70.6%. Taken together, these studies demonstrate the value of machine learning for identifying user interaction trends based on mobile device notifications.

### 2.4 Recognition and Emotional Impact of Notifications

Although machine learning has been used to identify user interactions with smartphones based on notifications, there has been little previous research using machine learning to predict the direct emotional impact of these notifications and interruptions. Many affective computing and HCI researchers have suggested various methods to sense and recognize human emotions (Kanjo, Al-Husain et al. 2015). Existing emotion recognition technologies can be divided into various categories, depending on what kinds of data are analyzed for recognizing human emotion: physiological signals, facial expressions, text or voice. Physiological emotion recognition shows acceptable performance but has some critical weaknesses that prevent its widespread use: they are obtrusive to users and need special equipment or devices (such as a skin conductance sensor, blood pressure monitor, or electrocardiography (ECG)). These devices are not only intrusive to users, but involve additional costs. Similarly, emotion recognition using facial expressions or speech limits their usage because the device needs to be positioned in front of the user's face and needs to continuously listen to the user's voice or record the user's face. This is not only not practical in the mobile context, but it raises various issues with regards to user acceptance and ethics.

While emotional recognition using physiological signals and facial/voice recognition has been extensively researched (Kanjo, (2017); Alhusain, (2013); Mawas, (2013); Alajmi, (2013); AlBarrak, (2013)), few studies have looked into the emotional responses to smartphone notifications. Many studies have found that notifications are predominantly linked to negative emotions. Pielot, Church et al.'s (2014) qualitative analysis revealed that using emails and social networks was correlated with feeling overwhelmed, stressed, interrupted and annoyed. Furthermore, when receiving more emails, participants were also more likely to report experiences where notifications kept them from doing something else or when they felt pressure to respond faster than they were able to. On the other hand, social notifications, despite their equally high volume, and social networks to a certain extent, caused more positive emotional responses. For example, it was found that receiving more social messages is significantly correlated with increased feelings of being connected with others. It is likely that this relates to the personal nature of messaging apps. Therefore, it would appear that the nature of notifications (work vs social) in this context rather than the amount of notifications may have an impact on users' emotional states, which will be assessed in the present study.

### 2.5 Impact on mental activity and affective states

Nomophobia is defined as fear of being without the mobile phone, resulting in discomfort, anxiety, and stress, and has been considered to be included in the most recent fifth edition of the Diagnostic and Statistical Manual for Mental Disorders (DSM-5) (Bragazzi & Del Puente, 2014). Psychologists recommend cognitive-behavioral psychotherapy combined with pharmacological intervention for the treatment of this potential disorder (Bragazzi & Del Puente, 2014).

Recent research findings have also shown that young adults who make particularly heavy use of mobile phones and computers run a greater risk of sleep disturbances, stress and symptoms of mental illness (Billieux et al., 2014). Thomée et al. (2011) conducted four studies assessing how the use of computers and mobile phones affects the mental health of young adults. Their findings have highlighted the need for moderation in the use of these technological devices. Their studies furthermore revealed that intensive use of mobile phones and computers can be linked to stress, sleep disorders and depressive symptoms. They also discovered that frequently using a computer or phone without breaks also increases the risk of stress, sleeping problems and depressive symptoms in women, whereas men who use computers intensively are more likely to develop sleeping problems. According to the mental health charity Young Minds (2016), problems highlighted as a consequence of excessive mobile phone use include the following: addiction, attention deficiency, attentiveness/or lack of it, depression, anxiety and stress, sleep disturbances and insomnia, as well as a lack of involvement in family life, suggesting excessive mobile phone use can be associated with a considerable array of difficulties.

## DATA COLLECTION SYSTEM

### 2.6 Software

The first stage of developing the classification system was the collection of notification data and affective state test responses. The data collection process is visualized in Figure 2. In this process, we gathered various notification data using the specifically designed mobile application 'NotiMind', which sits in the background and collects various system, message and social media notifications (see Figure 2). These notifications usually appear on the notification panel. NotiMind utilises the phone's Notification manager API and System Manager API in order to intercept notifications. These notifications are then logged and stamped with the time and date of the notification activity.

The application also records the following attributes (see Table 1):
1. Notification originators event name: e.g., email or social media client, such as WhatsApp and individual users or groups.
2. Event State in terms of what type of notification is being sent: e.g. screen event (Screen on, Screen-off and screen unlock), notification post (i.e. notification is received) and Notification removed (by the user).
3. Message content.
4. Event time and data.

NotiMind records the message body without requiring root privileges. However, users are required to manually enable notification access for NotiMind from their phone settings. All the data logged from the application are stored on a local SQLite database.

The application also collects self-reported affective states based on the Positive and Negative Affect Schedule (PANAS) model (Watson, Clark, & Tellegen, 1988). The PANAS model is based on the idea that it is possible to feel good and bad at the same time (Larsen, McGraw et al. 2001). Thus, PANAS tracks positive and negative affect separately. The PANAS contains adjectives to assess affect/mood states using differentiated terms (e.g. inspired, ashamed, and determined) with a general positive-negative index. The PANAS scale has good reliability, is sensitive to changes over time, and is considered one of the best measures of current mood (Watson, Clark, & Tellegen, 1988). To perform a measurement, the PANAS model uses a checklist to measure affect from various aspects. To reduce the burden on participants from completing measures of all the 60 PANAS elements, we assessed only 10 items based on a shortened PANAS version, the I-PANAS-SF (Thompson, 2007; see PANAS screenshot in Figure 2). All participants are asked to take the short PANAS test three times every day. The application sends a reminder every 8 hours to prompt users to take these tests. Also, users were encouraged to take the PANAS more often and whenever they can during the day to report their emotions.

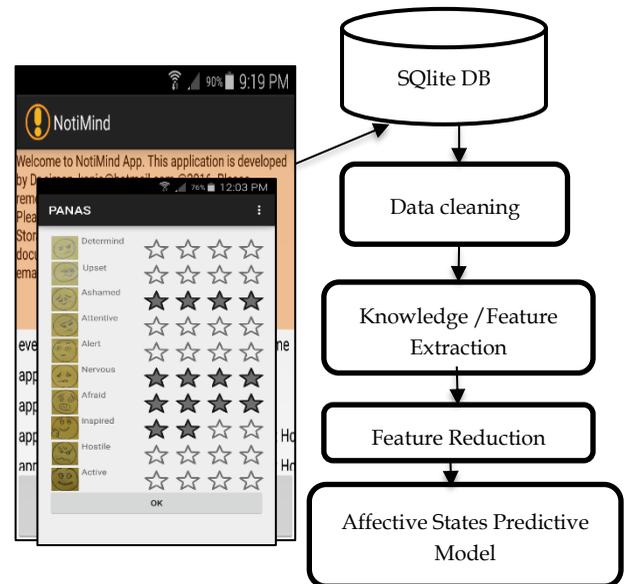

Fig 2. NotiMind system architecture and Android Platform

Data cleaning and features extraction was performed using Python, and our classifiers were performed using RapidMiner software (RapidMiner, 2016).

### 2.7 Participant Recruitment of the Participants

The study was launched in April 2016 over a five week period. The call for participation was advertised at Nottingham Trent University and the University of Kent's mailing lists, as well as on Twitter and other social media. The NotiMind application was not made available to download on the Google Play store for the following three reasons: 1) anyone can download apps on GooglePlay, even underage participants which violates our ethical approval agreement, 2) we cannot validate the purpose of data use; 3) no immediate benefit for the users is achieved, which meant a poor ranking might be given to the application by reviewers, which will undermine our future plan to release a consumer-facing version. Once participants have given their consent to participate in the study, they were asked to download the NotiMind app and enable Notification access on their mobile phones. They also received an instruction manual and email address in order to get in touch in case of any difficulties. The instruction manual pointed to where the log file was stored on the participant's Android phone (in our case it is a Sqlite file in .db format).

Our application triggered three silent alerts a day with popup notifications to remind the users of taking the PANAS test and to check that NotiMind is working as intended. Those who successfully completed the study were presented with a voucher as incentive for volunteering their time.

Fifty participants took part in the study (30 male and 20 female). All were existing Android users with an average age of 25. The study was approved by Nottingham Trent University's Ethics Committee.

During the data collection process, 832,776 notifications were collected. Many users stopped responding to the questionnaires after a few days and some did not respond at all. Therefore, we selected a subset of the data for the analysis and included data of 34 users who completed the data collection task. If no PANAS test score was recorded over 10 hours, then the data for this period were dismissed. Therefore, our final dataset comprises 534,346 notifications, and 3328 unique PANAS entries.

## 2.8 Response Rate

We first calculated the response rate of the overall PANAS affective measure input, confirming the reliability of user input into our system.

Although we asked users to report their mood at least three times a day, we expected some absence of self-report entry due to the extensive nature of our study. We calculated the daily response rate of our users and found that on average, 70% of users had three or more entries, which demonstrates the effectiveness of our self-report application. The affect distributions were fairly consistent during different times, seemingly unaffected by whether the entry was during work mornings, afternoons or evenings. The response rate, affect persistence, and distribution of the entered moods show NotiMind's efficacy in collecting a wide variety of data from the users, and provide insight into the design of NotiMind.

## 3 DATA PREPARATION

### 3.1 Basic definitions and Attributes

The NotiMind app generates five attributes, which are described in Table 1.

Due to the unstructured nature of these data, a considerable amount of cleaning, removing redundant data and reformatting was required. This step was followed by a feature extraction step (from Event-Name, and Event-Message columns).

Similar to prior work (Shirazi, Henze et al. 2014, Iqbal et al., 2010), we have derived a specific categorisation scheme based on messages and the applications they originated from. These new categories were then discussed with the users and finalised through further team discussion.

Also, following a similar approach to previous research (Shirazi, Henze et al. 2014), notifications were categorised as "work", if the message originated from email applications. Also we have found that "Group" messages names were often tagged with the symbol @, which made it easy to detect a "Group" category. Table2 shows the derived categories.

TABLE 1
DESCRIPTION OF ATTRIBUTES COLLECTED FROM THE NOTIMIND APP

| Attributes | Description |
| --- | --- |
| Date/Time | Time and Date of notification arrival |
| Event-Name | Including which application or internal mobile service initiated the message, phone number or name of a user, or a group ( e.g. com.android.systemui : Cable charging). |
| Event-State | With two possible values, either Notification is posted or removed from the notification panel |
| Event-Message | Content of the message |
| Type | Notification event or screen event, such as screen locked screen On or Off |

TABLE 2
LIST OF CATEGORIES DERIVED FOR THE AFFECT STATE PREDICTION

| Categories | Description |
| --- | --- |
| Group | Detected when the '@' sign is present in the Event-Name. e.g. whatsapp : William @ Friendship-Group |
| Work | Detected when certain email tags are present in the Message. Also messages received from LinkedIn are considered as work related. |
| System | Detected when the following keywords are present in the Message or Event-Name, including: 'Updating', 'WIFI', or 'USB'. |
| Emoji Description | Emojis detected in the text are decoded to its description according to the Emoji/Unicode table*. |
| Emoji Count | The number of Emojis in the message. |
| Video Presence | Video presence in the message is represented by the 'Camera Recording' Emoji (🎥 Unicode: U+1F3A5). |
| Video Length | Length of the video is recorded from the value accompanying the 'Camera Recording' Emoji |
| Image Presence | Image presence in message is represented by the 'Camera' Emoji (📷, Unicode: U+1F4F7). |
| Message Length | Length of the message. |
| Multiple Messages | Multiple messages notifications don't show actual messages, however, they indicate that the user is receiving many number messages simultaneously. |

### 3.2 PANAS Data distribution

We considered the PANAS as a two-dimensional vector with 20 possible values for PA (Positive Affect) and 20 possible values for NA (Negative Affect).

PA is the sum of PANAS items 1, 4, 5, 8, 10 (positive metrics (PM)):

*PM= <Determined, Attentive, Alert, and Active>*

$$PA = \sum_{i}^{t} PM$$

NA is the sum of PANAS items 2, 3, 6, 7, 9 (negative metrics (NM)):

*NM=<Inspired, Ashamed, Nervous, Upset, Afraid Hostile>*

$$NA = \sum_{i}^{t} NM$$

In the current study, following Koydemir et al.'s (2013) and Mukolo and Wallston's (2012) method, the affect balance score was computed as overall indicator of affective well-being by subtracting the negative affect score from the positive affect score (*PA-NA)*. The affect balance (we will refer to it as the PANAS score) has the range *L*, ranging from -20 to 20. We discretize this range into 3 classes: *-1* (negative), *0* (neutral), and *1* (positive). To perform data discretization, we adopted a recursive partitioning based on entropy of the PANAS score distribution. Intuitively the entropy measures the amount of randomness of a source producing random items. Using this approach, an interval is split at a point that results in minimum entropy. Formally, let $p_i$ be the empirical probability of observing the label $y_i$ on sequence.

*I*, i.e., the ratio of labels $y_i$ to all labels in the sequence *I*. Then the entropy of the label distribution on *I* is defined as:

$$\text{Entropy}(D_i) = -\sum_{i=1}^{m} p_I(y_i) Log_2(1/p_I(y_i))$$

where the sum is over-all labels.

Discretization ranges are determined by selecting the cut point for which Entropy is minimal amongst all the candidate cut points (Fayyad & Iran, 1993).

We also analyzed the distribution of PANAS scores that users entered into the system. As we expected, neutral affective states occupy a significant percentage of our dataset.

Users generally reported positive affective states rather than very negative states. 68% of total PANAS affective states reported by participants were positive, while 32% of total reported PANAS affective states were negative. The maximum number of positive states reported was +18, and the minimum negative states reported was -5, with a mean µ=4.78 and a Standard Deviation of σ=5.08.

13% represents the highest PANAS entry (mood state *Active*), whilst 4% represents the lowest PANAS entry (mood state *Ashamed*) as shown in Table 3.

TABLE 3

PANAS Scores Frequencies

| **PANAS Individual Metrics** | |
|---|---|
| Determined | 6% |
| Upset | 9% |
| Ashamed | 4% |
| Attentive | 8% |
| Alert | 9% |
| Nervous | 5% |
| Afraid | 5% |
| Inspired | 9% |
| Hostile | 5% |
| Active | 13% |
| PANAS Positive and Negative | |
| PA | 68% |
| NA | 32% |

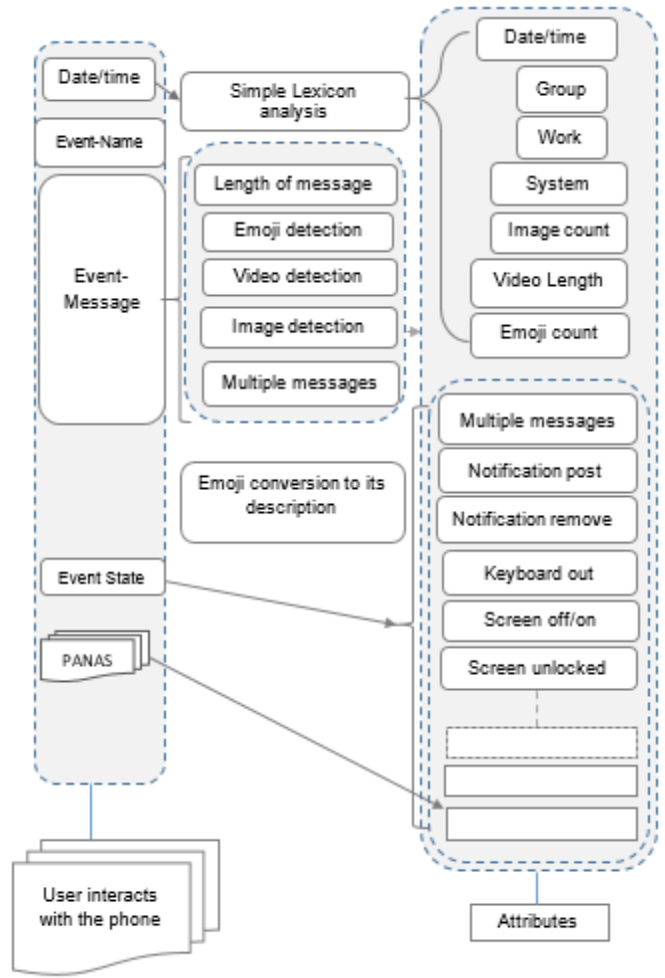

Fig. 3. Feature extraction Dataflow

In addition, Figure 3 shows the whole life cycle of data collection and categorisation including the PANAS self-report process.

Figure 4 shows the count of PANAS annotations per user. Although all users were reminded to take the test three times a day (at the same interval), the level of motivation to annotate varied between users.

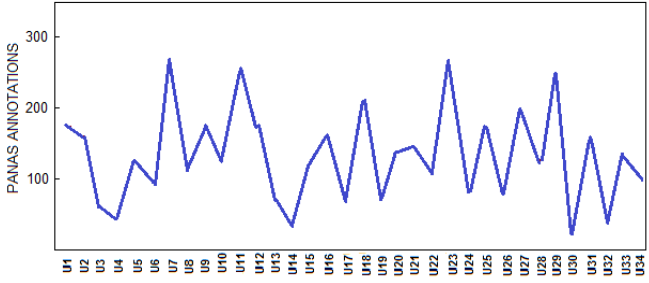

Fig 4. Number of PANAS annotations per user (STD=±63).

### 3.3 Data Segmentation

In total, we utilized 534,346 valid notifications from 34 users. Based on the derived categories above, we segmented the data based on PANAS scores (one segment per PANAS label) and then we extracted our features based on the following process:

In each segment, the rate of occurrence (frequency) for each metric (category) was calculated as follows:

Let $t_1$ be the time of PANAS test, $t_2$ the time of the following PANAS test. Then the sum of Post Notification P during ($t_i$, $t_i+1$):

$$P(t_i, t_{i+1}) = \sum_i^t P_i$$

Then we calculated the percentage Post Notifications $P$ for one segment based on the following formula:

$$P_A = (\sum_i^t P(t_i, t_{i+1}) * 100) / (\sum_i^t N(t_i, t_{i+1}))$$

where $N(t_i, t_{i+1})$ is the total number of notifications in the segment.

Similarly, R($t_i$,$t_i+1$), O($t_i$,$t_i+1$), F($t_i$,$t_i+1$), U($t_i$,$t_i+1$), and K($t_i$,$t_i+1$), M($t_i$,$t_i+1$) and G($t_i$,$t_i+1$), are the sums for Removed, Screen-On, Screen-off, Unlock, Keyboard-Out, Multiple and Group states over ($t_i$,$t_i+1$) period, respectively. The corresponding percentages are $R_A$, $O_A$, $F_A$, $U_A$, and $K_A$ and were calculated similarly.

In addition, the following percentages, Emoji count $E_A$, Work Notification, Rate $W_A$, Group Notification $G_A$, Multi Notification $M_A$ and System Notification $S_A$ were calculated as follows:

$$E_A = (\sum_i^t R(t_i, t_{i+1}) * 100) / (\sum_i^t P(t_i, t_{i+1}) - \sum_i^t M(t_i, t_{i+1}))$$

We excluded the Multi-Notifications from the Post notifications count since these notifications do not have Emojis in them.

$$W_A = (\sum_i^t R(t_i, t_{i+1}) * 100) / (\sum_i^t P(t_i, t_{i+1}) - \sum_i^t S(t_i, t_{i+1}))$$

Similarly, we excluded the System notifications from the total post notification count in order to calculate the work percentage.

$$S_A = (\sum_i^t R(t_i, t_{i+1}) * 100) / \sum_i^t P(t_i, t_{i+1})$$

$$M_A = (\sum_i^t M(t_i, t_{i+1}) * 100) / \sum_i^t P(t_i, t_{i+1})$$

$$G_A = (\sum_i^t G(t_i, t_{i+1}) * 100) / \sum_i^t P(t_i, t_{i+1})$$

The segmentation process resulted in *3328* segmented instances (rows) of labelled data. Table 4 lists all the categories along with their percentages based on the overall data.

TABLE 4
MAIN EXTRACTED FEATURES, WITH THE OVERALL PERCENTAGE.

| Symbol | Extracted Features | Percentage of overall data (%) |
|---|---|---|
| P | Notification Post | 31 |
| R | Notification Removed | 5 |
| O | Screen On | 15 |
| F | Screen Off | 15 |
| U | Screen Unlock | 10 |
| K | Keyboard-Out | 20 |
| S | System Notification | 6 |
| M | Multiple Notification | 17 |
| E | Emoji Count | 9 |
| G | Group Notifications | 13 |
| W | Work Notifications | 7 |
| $t_i$ | Time and Date of the event. | |
| V | The phone's ringer mode when notification arrived (Volume) | |

## 4 CORRELATION ANALYSIS

By examining the statistical significance of our variables (i.e., features), we determined the relation between our features and the affective measure as well as interdependency between these variables. The highest positive correlations were observed between Keyboard-Out and PANAS score (r = 0.46, p< 0.001). Figure 5 depicts the lack of negative affective states when high instances of keyboard states were present. This suggests that users experienced positive emotions when interacting with other users (while the user is typing a message).

Our data showed that only 10% of Keyboard-out data segments (when Keyboard-Out=1) are linked to negative PANAS.

On the other hand, the data demonstrate more negative emotions are present when the users constantly receive and remove messages without direct engagement with other users as shown in Figure 6. Our data showed that 89% of the segments with high Post and Remove Notifications rates are associated with negative PANAS scores.

A negative correlation was found between work messages and PANAS scores (r = −0.38, p < 0.001). This indicates that people like to engage with social messages and family and friends, and become stressed when interacting with work-related messages.

There was a noticeable positive correlation between multiple messages notifications (e.g., five messages from WhatsApp notifications which did not reveal the identity of the senders) and PANAS score (r = 0.22, p < 0.001).

Emoji Count was noticeably correlated with high positive affective measures (r =0.35). A closer look identified a statistically significant correlation between Emojis count and positive PANAS scores (p <0.01).

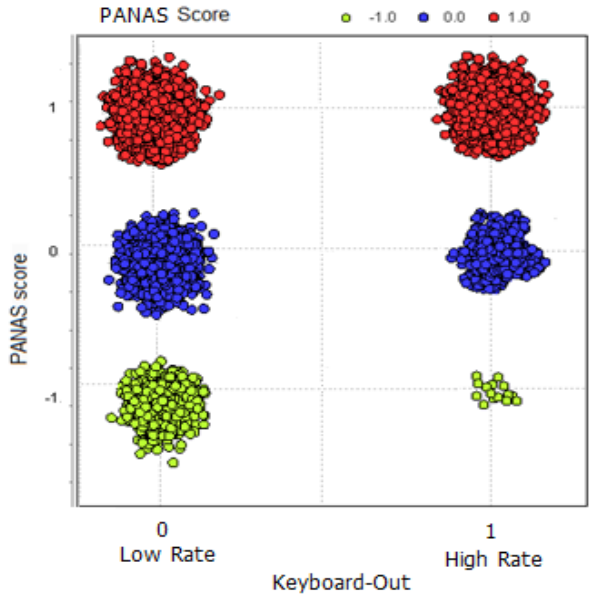

Fig. 5. Shows a correlation between low rate of Keyboard-Out states occurrences and negative affective measures

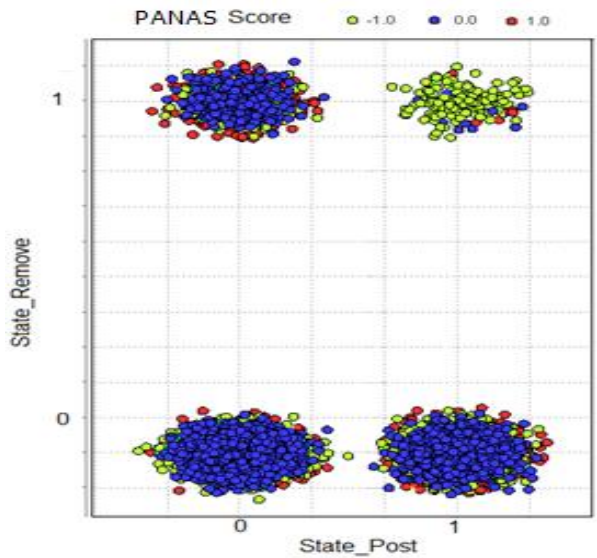

Fig. 6. Shows that high rate of Post and Remove states correlates with high negative affective measures.

## 5 PREDICTIVE MODEL

In order to develop the system for the identification of affective states from notifications and mobile usage behavior, we developed a predictive model. For these tasks, we utilized a popular machine learning software toolkit named RapidMiner Data Mining software v7.1 (RapidMiner, 2016) and also checked the results on the Machine Learning Open Source software WEKA 3.8 (Weka, 2016). As mentioned above, emotional states were defined by three classes in the classification task: -1, 0 and 1, for negative, neutral and positive, respectively.

To build the model, we tested the levels of significance of the features in relation to the PANAS scores and checked the response of the PANAS scores for any interdependency between the variables based on the correlation matrix.

We checked the pairwise correlations between features and PANAS score on the whole dataset. Based on the result of features evaluation, we finally selected nine features, which have strong correlations with PANAS scores to build an inference model (i.e., feature selection step). Selected features were: Keyboard-Out, Emoji-Count, Remove, Work, Post, Group, Multi, Screen-On and Unlock.

TABLE 5
PAIRWISE CORRELATIONS BETWEEN FEATURES AND PANAS SCORE.

| Features | r |
|---|---|
| Keyboard-Out | 0.46 |
| Emoji-Count | 0.35 |
| Multi | 0.22 |
| Post | 0.13 |
| Screen-On | 0.09 |
| Remove | 0.08 |
| Unlock | -0.07 |
| Group | -0.08 |
| Work | -0.35 |

As shown in table 5, Keyboard-Out had the highest positive correlation with PANAS score, while Work had a high negative correlation. We kept Screen-On state and removed Screen-off since highly correlated attributes could lead to a multicollinearity problem.

In order to configure Rapid Miner to build the predictive model, the selected features were then normalized and converted into polynomial data as shown in Figure 7, which represents the main building blocks of our predictive model using RapidMiner.

Then we set a special role which identified a label (i.e. the PANAS scale), which must be predicted for new examples that were not yet characterized in such a manner. Setting the label was a RapidMiner preparatory step to feed the data to the classifier to build a predictive model of the notification data.

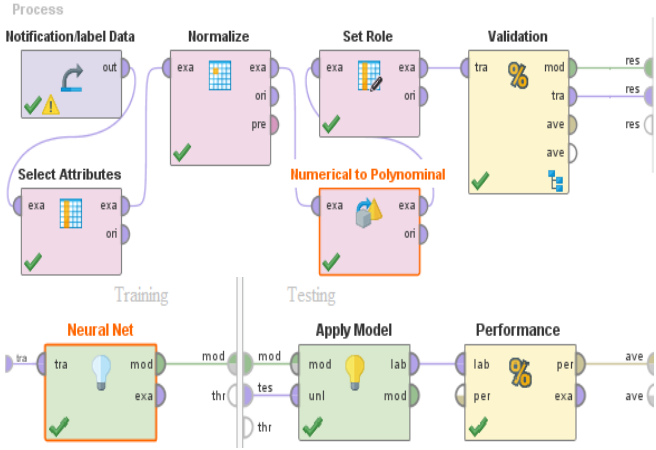

Fig. 8. Rapid Miner design of ANN predictive model

The performances of three supervised machine learning algorithms were tested to classify user interaction segments into three PANAS classes. A feed-forward Neural Network (ANNs) with a hidden layer, and radial basis function-Support Vector Machine (rbf-SVM) (Chang, Hsieh et al. 2010), and a Logistic Regression (LR) analysis (Cox 1958) were performed. We chose these algorithms in order to evaluate how discriminative (SVM), probabilistic (LR) and neural network (*ANN*) (McCulloch & Pitts 1943) algorithms work on our dataset.

A neural network system (ANN) for recognition is defined by a set of input neurons (nodes) which can be activated by the information of the intended object to be classified. The input can be either raw data, or pre-processed data from the samples. In our case, we have pre-processed our data by building a feature vector. The feed-forward neural network was trained by a back propagation algorithm (single layer). An artificial neural network (ANN) is a mathematical model or computational model that is inspired by the structure and functional aspects of biological neural networks.

Our ANN was set with one hidden layer, 500 learning cycles and a 0.3 learning rate and a momentum of 0.2. The momentum simply added a fraction of the previous weight update to the current one. This prevented local maxima and smoothed optimization directions. It indicated whether the learning rate should be decreased during learning.

The averaged performance of each classifier was assessed via a multiple-run k-fold (nested) stratified cross-validation. In our study, we adopted fifteen and ten folds. The inner loop of the cross-validation aimed to perform model selection. To quantify the performance of the classification models, we used the F-Measure, the harmonic mean of precision and recall, as our primary evaluation metric. F-Measure is calculated as follows:

$$F - Measure = 2 \cdot \frac{Precision \cdot Recall}{Precision + Recall}$$

Figure 7 shows the performance of the three classifiers across the two cross-validation methods.

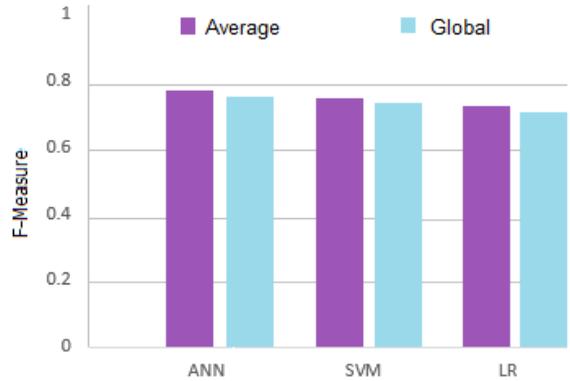

Fig.8. F-Measure of different Classification Models: ANN, SVM and LR.

TABLE 6
F-MEASURE OF THE MODELS

|  | ANN | SVM | LR |
|---|---|---|---|
| **Average** | 0.784 | 0.762 | 0.741 |
| **STD** | ±0:09 | ±0:15 | ±0:16 |
| **Global** | 0.753 | 0.752 | 0.721 |

We employed 15-Fold Cross-Validations to evaluate the within-subject models, and Leave-One-Out for the global one. F-measures were calculated to evaluate the goodness of the classification model among the three classification methods (ANN, SVM and LR). When a comparative study comprises $n$ classifiers, $c = \frac{n(n-1)}{2}$ pairwise comparisons are possible.

As shown in Figure 8 and Table 6, we observed that *ANN* worked significantly better than SVM (p-value=0:007) using the within-the-subject model, and SVM was significantly better than LR (p-value=0:009). ANN was significantly better than LR (p-value=0:002).

By applying Bonferroni Adjustment, we obtained the critical value as follows: $\alpha$ =0.0.5/3=0.016. The statistical power was then calculated based on $\alpha$. This means the null hypothesis of equal performance was rejected for all the comparative tests (p-value<0.016).

Similarly, we obtained similar results when applying comparative pairwise tests on the Global models (p-values were 0.04, 0.012 and 0.003, respectively), and the null hypothesis of equal performance was also rejected (p-value<0.016).

The data from our first field study with participants showed that it is possible to create a machine-learning model to automatically predict when people are experiencing positive, neutral or negative affective states from smartphone notification activities with F-measures between 74-78% for the within-subject model, and between 72-76% for the global model.

In the next section, we will demonstrate how we can improve the accuracy level of the model performance by conducting another experiment when the user is in a less disruptive environment.

## 6 DISCUSSION

The main focus of this work was to explore whether notification and screen interaction information extracted from smartphones can be a good predictor of users' affective states. We took a systematic approach to model the dependency of mobile notifications and screen interaction patterns and users' emotions. We demonstrated that our model can predict users' affect state with good performance, as shown in the previous section. Furthermore, our initial hypothesis was that all interactions with mobile phone notifications have a negative impact on affective states of users. Our findings indicate that users feel positive when engaged directly with other humans using social media applications.

One possible interpretation of this finding is that people are more willing to share their positive emotional states, while those experiencing negative emotions are reluctant to broadcast how they feel or share information in general. We discussed these results with some of the participants after sharing our findings, and our data interpretation was confirmed.

On the other hand, notifications of non-social messages (e.g., WIFI availability and system updates) have the most negative impact on users. Users usually get frustrated when they receive a notification which is not sent by a human.

Positive affective states further correlated with multi-messages notifications (e.g., 5 messages from WhatsApp) which did not show the identity of the senders directly, but created a sense of belonging and feeling more connected. As expected, work-related notifications had a negative impact on affective states, especially when they arrived in bulks.

The presence of Emojis in notification messages influenced the sentimental value of the message and hence correlated with positive affective states of users who were receiving them. This is not surprising since we have found recently a significant relationship between the number of Emojis and the overall positive affective measure in a relevant project (Tauch & Kanjo, 2016). Emoji characters may seem trivial, but they are becoming the world's fastest growing language in all forms of communications, text messages, posts on social media, chatting applications and emails (Kralj Novak, Smailović et al. 2015). This may be due to a richer set of graphical representations of facial expressions in comparison to text only, which may lead to improved reader comprehension of the emotional message content, and a visual representation of animals, food, activities, etc.

### 6.1 Limitations

Our evaluation results presented in Section 5 illustrate the promising nature of affective states sensing based on interaction with messages and notifications, despite noisy labels and variation in user input. Our future work is required to more thoroughly evaluate our models on larger datasets.

We outline the limitations of our work subsequently so future work can address these.

**Incentivized Public Deployment:** Currently our NotiMind application and our call for participation have managed to attract a reasonably high level of user participation (mainly from within university contexts). Although we originally recruited 50 users, only 34 users' data have provided PANAS labels. Future work will look at improving user participation and enhancing various user interface features and performance to enable a robust, continuous and flawless collection phase.

**Robustness of Self-reports:** A remaining challenge in training data from sensing systems is the collection of accurate subjective labels from users. Similarly, annotating affective responses can be imprecise, since users may deviate in expressing their own emotions, or mapping them differently on a scale.

**Additional Affective states cues:** With the advancement of the off-the-shelf wearable devices that can monitor health and physiological change, we can easily combine our approach with recent work of others exploring alternative data sources for mental health inference, such as sensor data. This could account for other co-founder variables that may influence mood and affective states.

Finally, we will look at new techniques to extract features that are not limited to a specific window size. Instead, we will explore the potential of adaptive data segmentation and recurrent networks which is one of the staples of deep learning that allow to contextualize and learn the temporal dynamics of sequential data across long periods of time.

## 7 POTENTIAL APPLICATIONS

In summary, the present research paves the foundation for future studies looking into predicting emotions based on smartphone user responses to notifications and activities. In the long term, research in this direction will have a significant impact, as outlined in the following:

1. **Impact on society**: Emotion detection could help improve behavior prediction. For example, people who discuss dangerous and violent acts, but seem to be happy and emotionally stable, might be less of a threat to society or themselves than people who do not discuss these subjects in their messages, but their overall negative emotions might make them more of a concern. This has implications for national security and cybersecurity.

2. **Intervention:** The output of our predictive model could be reflected on users' profiles (e.g., social media applications, such as WhatsApp and Facebook) which could prompt positive interventions, such as fewer system notifications and more messages from close friends when the user is feeling down. This may be particularly relevant for younger users who may be overly preoccupied with their messages.

3. **Entertainment features on the phone:** Knowledge about users' affective states can be abstracted in a decorative character (i.e. Emoji style agent) which can be displayed on their phone. Illustrating the user's emotional state may allow the user to insert this emotion into messages or social media posts.

In exploring these possible applications, we need to consider the ethical implications of monitoring people's smartphone activities as this can be seen as an intrusion of privacy. Although this form of monitoring is physically less intrusive than physiological signals (i.e., not requiring the users to wear specialist equipment), people may be concerned about their private text messages being "read" by the system, and thus may be unwilling to use it despite the potential benefits.

## 8 CONCLUSIONS AND FUTURE WORK

In this paper, we proposed an unobtrusive emotion recognition approach which exploits phone notifications as affective states sensor. The mobile application *NotiMind* was developed, which logs and samples both mobile notification interaction activities along with PANAS affective measures. Additionally, we proposed a machine learning method to automatically infer users' affective states from the collected mobile data: mobile notification and screen interaction (e.g., keyboard-out, Post and Remove), notification style (e.g., work, health or social) and message context (e.g., message length, presence of Emojis or images). In the user study, we gathered 534,346 notifications which formed the base of our training dataset including nine of 20 selected features. We built three classifiers and they showed good classification performance against three classes: positive, neutral and negative.

We identified a direct relationship between different notification interaction states and user's emotions. We also validated some past psychological studies which have suggested that mobile phone use and constant notifications and interruption can impact users' mental health (Billieux et al., 2015; Lopez-Fernandez et al., 2017).

The proposed technology can contribute to creating positive emotions for users through automatic recognition and sharing of their emotions. Suggestions for future research include improving the accuracy of classifications by employing more features associated with users' emotions. For instance, we will look at the sentimental contents of the messages. In addition, we intend to work on a real-time emotion analysis and prediction application which acts as a background service to keep users informed of their current mental states. This will help users to refrain from excessive engagement with their mobile phones and with disruptive applications by switching off unnecessary notifications. The resultant emotive indicators can be tagged on to social media profiles which may allow users to be aware of their friends' current mood and wellbeing.

## ACKNOWLEDGMENT

This research was supported by the British Academy and the Leverhulme Trust.

## REFERENCES

Adamczyk, P. D. and B. P. Bailey (2004). If not now, when?: the effects of interruption at different moments within task execution. Proceedings of the SIGCHI Conference on Human Factors in Computing Systems. Vienna, Austria, ACM**:** 271-278.

Alajmi, N. , Kanjo E., Mawass, N.E., Chamberlain, A: Shopmobia: an emotion-based shop rating system, Affective Computing and Intelligent Interaction (ACII), (2013), pp. 745–750

Al-Barrak, L, E. Kanjo, NeuroPlace: making sense of a place, Augmented Human (2013), pp. 186–189

Al-Husain,L, E. Kanjo, A. Chamberlain, Sense of Space: Mapping Physiological Emotion Response in Urban Space, UbiComp (Adjunct Publication) (2013), pp. 1321–1324

Avrahami, D. and S. E. Hudson (2006). Responsiveness in instant messaging: predictive models supporting inter-personal communication. Proceedings of the SIGCHI Conference on Human Factors in Computing Systems. Montréal, Québec, Canada, ACM**:** 731-740.

Baethge, A. and T. Rigotti (2013). "Interruptions to workflow: Their relationship with irritation and satisfaction with performance, and the mediating roles of time pressure and mental demands." Work & Stress **27**(1): 43-63.

Bayer, J. B. and S. W. Campbell (2012). "Texting while driving on automatic: Considering the frequency-independent side of habit." Computers in Human Behavior **28**(6): 2083-2090.

Begole, J. B., N. E. Matsakis, et al. (2004). Lilsys: Sensing Unavailability. Proceedings of the 2004 ACM conference on Computer supported cooperative work. Chicago, Illinois, USA, ACM**:** 511-514.

Billieux, J., Maurage, P., Lopez-Fernandez, O., Kuss, D. J., & Griffiths, M. D. (2015). Can disordered mobile phone use be considered a behavioral addiction? An update on current evidence and a comprehensive model for future research. Current Addiction Reports, **2**(2) 156-162. DOI: 10.1007/s40429-015-0054-y

Bragazzi, N. L., & Del Puente, G. (2014). A proposal for including nomophobia in the new DSM-V. Psychology Research and Behavior Management, **7**, 155-160. doi: 10.2147/prbm.s41386

Chamberlain, A, M. Paxton, K. Glover, M. Flintham, D. Price, C. Greenhalgh, S. Benford, P. Tolmie, E. Kanjo, A. Gower, A. Gower, D. Woodgate Danaë Emma Beckford Stanton Fraser: understanding mass participatory pervasive computing systems for environmental campaigns, Pers. Ubiquitous Comput., 18 (7) (2014), pp. 1775–1792

Chang, Y.-J. and J. C. Tang (2015). Investigating Mobile Users' Ringer Mode Usage and Attentiveness and Responsiveness to Communication. Proceedings of the 17th International Conference on Human-Computer Interaction with Mobile Devices and Services. Copenhagen, Denmark, ACM**:** 6-15.

Chang, Y.-W., C.-J. Hsieh, et al. (2010). "Training and Testing Low-degree Polynomial Data Mappings via Linear SVM." J. Mach. Learn. Res. **11**: 1471-1490.

M. Ciman; K. Wac, "Individuals' stress assessment using human-smartphone interaction analysis," in IEEE Transactions on Affective Computing , vol.PP, no.99, pp.1-1 doi: 10.1109/TAFFC.2016.2592504.

Cox, D. R. (1958). "The Regression Analysis of Binary Sequences." Journal of the Royal Statistical Society. Series B (Methodological) **20**(2): 215-242.

Czerwinski, M., E. Horvitz, et al. (2004). A diary study of task switching and interruptions. Proceedings of the SIGCHI Conference on Human Factors in Computing Systems. Vienna, Austria, ACM**:** 175-182.

Fogarty, J., S. E. Hudson, et al. (2005). "Pred (Watson, Clark, & Tellegen, 1988)icting human interruptibility with sensors." ACM Trans. Comput.-Hum. Interact. **12**(1): 119-146.

Horvitz, E. C. M. C. E. (2001). Notification, Disruption, and Memory: Effects of Messaging Interruptions on Memory and Performance, IOS Press.

Hudson, S., J. Fogarty, et al. (2003). Predicting human interruptibility with sensors: a Wizard of Oz feasibility study. Proceedings of the SIGCHI Conference on Human Factors in Computing Systems. Ft. Lauderdale, Florida, USA, ACM**:** 257-264.

Ichikawa, F., J. Chipchase, et al. (2005). Where's The Phone? A Study of Mobile Phone Location in Public Spaces. 2005 2nd Asia Pacific Conference on Mobile Technology, Applications and Systems.

Iqbal, S. T. and E. Horvitz (2010). Notifications and awareness: a field study of alert usage and preferences. Proceedings of the 2010 ACM conference on Computer supported cooperative work. Savannah, Georgia, USA, ACM: 27-30.

Kanjo, E., L. Al-Husain, et al. (2015). "Emotions in context: examining pervasive affective sensing systems, applications, and analyses." Personal and Ubiquitous Computing 19(7): 1197-1212.

Kanjo, E, Eman M.G. Younis, Nasser Sherkat, Towards unravelling the relationship between on-body, environmental and emotion data using sensor information fusion approach, Information Fusion, Volume 40, March 2018, Pages 18-31, ISSN 1566-2535, https://doi.org/10.1016/j.inffus.2017.05.005.

Koydemir, S., Faruk, Ö., Schütz, Ş. A., & Tipandjan, A. (2013). Differences in How Trait Emotional Intelligence Predicts Life Satisfaction: The Role of Affect Balance Versus Social Support in India and *Germany.* Journal of Happiness Studies. doi:10.1007/s10902-011-9315-1.

Kralj Novak, P., J. Smailović, et al. (2015). "Sentiment of Emojis." PLoS ONE **10**(12): e0144296.

Kushlev, K. and E. W. Dunn (2015). "Checking email less frequently reduces stress." Computers in Human Behavior **43**: 220-228.

Kushlev, K., J. Proulx, et al. (2016). "Silence Your Phones": Smartphone Notifications Increase Inattention and Hyperactivity Symptoms. Proceedings of the 2016 CHI Conference on Human Factors in Computing Systems. Santa Clara, California, USA, ACM**:** 1011-1020.

Larsen, J. T., A. P. McGraw, et al. (2001). "Can people feel happy and sad at the same time?" Journal of personality and social psychology **81**(4): 684-696.

Lopez-Fernandez, O., Kuss, D. J., Romo, L., Morvan, Y., Kern, L., Graziani, P., Rousseau, A., Rumpf, H.J., Bischof, A., Gässler, A. K., Schimmenti, A., Passanisi, A., Männikkö, N., Kääriänen, M., Demetrovics, Z., Király, O., Chóliz, M., Zacarés, J. J., Serra, E., Griffiths, M. D., Pontes, H. M., Lelonek-Kuleta, B., Chwaszcz, J., Zullino, D., Rochat, L., Achab, S., & Billieux, J. (2016). Self-reported dependence on mobile phones in young adults: A European cross-cultural empirical survey. *Journal of Behavioral Addictions*, in press.

Mark, G., S. Voida, et al. (2012). "A pace not dictated by electrons": an empirical study of work without email. Proceedings of the SIGCHI Conference on Human Factors in Computing Systems. Austin, Texas, USA, ACM**:** 555-564.

McCulloch, W. S. and W. Pitts (1943). "A logical calculus of the ideas immanent in nervous activity." The bulletin of mathematical biophysics **5**(4): 115-133.

Mawass, N.E., E. Kanjo, A Supermarket Stress Map, UbiComp (Adjunct Publication) (2013), pp. 1043–1046

Oulasvirta, A., T. Rattenbury, et al. (2012). "Habits make smartphone use more pervasive." Personal and Ubiquitous Computing 16(1): 105-114.

Pielot, M., K. Church, et al. (2014). An in-situ study of mobile phone notifications. Proceedings of the 16th international conference on Human-computer interaction with mobile devices & services. Toronto, ON, Canada, ACM**:** 233-242.

Pielot, M., T. Dingler, et al. (2015). When attention is not scarce - detecting boredom from mobile phone usage. Proceedings of the 2015 ACM International Joint Conference on Pervasive and Ubiquitous Computing. Osaka, Japan, ACM**:** 825-836.

Pielot, M., R. d. Oliveira, et al. (2014). Didn't you see my message?: predicting attentiveness to mobile instant messages. Proceedings of the 32nd annual ACM conference on Human factors in computing systems. Toronto, Ontario, Canada, ACM: 3319-3328.

Pielot, M. and L. Rello (2015). The Do Not Disturb Challenge: A Day Without Notifications. Proceedings of the 33rd Annual ACM Conference Extended Abstracts on Human Factors in Computing Systems. Seoul, Republic of Korea, ACM: 1761-1766.

Przybylski, A. K., Murayama, K., DeHaan, C. R., & Gladwell, V. (2013). Motivational, emotional, and behavioral correlates of fear of missing out. Computers in Human Behavior*,* 29(4), 1841-1848. doi: http://dx.doi.org/10.1016/j.chb.2013.02.014.

RapidMiner. (2016, 08 19). *https://rapidminer.com*. Retrieved from https://rapidminer.com.

Rosenthal, S., A. K. Dey, et al. (2011). Using Decision-Theoretic Experience Sampling to Build Personalized Mobile Phone Interruption Models. Pervasive Computing: 9th International Conference, Pervasive 2011, San Francisco, USA, June 12-15, 2011. Proceedings. K. Lyons, J. Hightower and E. M. Huang. Berlin, Heidelberg, Springer Berlin Heidelberg: 170-187.

Shirazi, A. S., N. Henze, et al. (2014). Large-scale assessment of mobile notifications. Proceedings of the 32nd annual ACM conference on Human factors in computing systems. Toronto, Ontario, Canada, ACM: 3055-3064.

Tauch, C., & Kanjo, E. ( September 12-16, 2016). The roles of Emojis in Mobile Phone Notifications. Ubicomp/ISWC'16 Adjunct, Heidelberg, Germany, ACM.

Thomée, S., Härenstam, A., & Hagberg, M. (2011). Mobile phone use  and stress, sleep disturbances, and symptoms of depression among young adults - A prospective cohort study. BMC Public Health, 11(1), 1-11. doi: 10.1186/1471-2458-11-66

Thompson, E. R. (2007). "Development and Validation of an Internationally Reliable Short-Form of the Positive and Negative Affect Schedule (PANAS)." Journal of Cross-Cultural Psychology 38(2): 227-242.

Usama M. Fayyad, K. B. (1993). *Multi-interval discretization of continuousvalued attributes for classification learning.* : Thirteenth International Joint Conference on Articial Intelligence.

Watson, D., Clark, L. A., & Tellegen, A. (1988). Development and validation of brief measures of positive and negative affect: The PANAS scales (Vol. 54(6)). Journal of Personality and Social Psychology. doi:0022-3514.54.6.1063.

Weka. (2016, 08 19). *Weka 3: Data Mining Software in Java*. Retrieved 2016, from Weka: http://www.cs.waikato.ac.nz/

Westermann, T., S. M, et al. (2015). Assessing the Relationship between Technical Affinity, Stress and Notifications on Smartphones. Proceedings of the 17th International Conference on Human-Computer Interaction with Mobile Devices and Services Adjunct. Copenhagen, Denmark, ACM: 652-659.

Wiese, J., T. S. Saponas, et al. (2013). Phoneprioception: enabling mobile phones to infer where they are kept. Proceedings of the SIGCHI Conference on Human Factors in Computing Systems. Paris, France, ACM: 2157-2166.

Yoon, S., S.-s. Lee, et al. (2014). Understanding notification stress of smartphone messenger app. Proceedings of the extended abstracts of the 32nd annual ACM conference on Human factors in computing systems. Toronto, Ontario, Canada, ACM: 1735-1740.

Young Minds. (2016). Worried about your child? Internet and mobile phone use. Retrieved 16.08.2016, from http://www.youngminds.org.uk/for_parents/worried_about_your_child/internet_mobile